\documentclass[aps,prl,superscriptaddress,amsmath,amssymb,twocolumn,footinbib]{revtex4-1}

\usepackage[T1]{fontenc}
\usepackage[utf8]{inputenc}
\usepackage[caption=false]{subfig}
\usepackage{graphicx} 
\usepackage{epstopdf}
\usepackage{array}
\usepackage{verbatim}
\usepackage{amsmath,amsfonts,amssymb,amscd,mathtools,bbm}
\usepackage{amsthm}
\usepackage{tabularx}
\usepackage{stmaryrd}
\usepackage{enumerate}
\usepackage[ruled]{algorithm2e}
\usepackage{wasysym}
\usepackage{braket}
\usepackage[dvipsnames]{xcolor}
\usepackage{mathrsfs}
\usepackage{microtype}
\usepackage{hyperref}
\hypersetup{
    bookmarksnumbered=true, 
    breaklinks=true,
    unicode=false, 
    pdfstartview={FitH}, 
    pdftitle={}, 
    pdfnewwindow=true, 
    colorlinks=true, 
    allcolors=MidnightBlue!70!black!15!blue
}
\usepackage{newtxtext,newtxmath}
\usepackage[normalem]{ulem}
\usepackage{url}

\theoremstyle{plain}
\newtheorem{thm}{Theorem}

\newtheorem{lem}[thm]{Lemma}

\theoremstyle{definition}

\let\nc\newcommand

\nc{\R}{R} 
\nc{\Rl}{\underline{R}} 
\nc{\Rs}{R^{s}} 
\nc{\Rsl}{\underline{R}^{s}} 

\newcommand{\eq}[1]{(\hyperref[eq:#1]{\ref*{eq:#1}})}

\renewcommand{\sec}[1]{\hyperref[sec:#1]{Section~\ref*{sec:#1}}}
\newcommand{\thrm}[1]{\hyperref[thrm:#1]{Theorem~\ref*{thrm:#1}}}
\newcommand{\lemm}[1]{\hyperref[lemm:#1]{Lemma~\ref*{lemm:#1}}}
\newcommand{\prop}[1]{\hyperref[prop:#1]{Proposition~\ref*{prop:#1}}}
\newcommand{\corr}[1]{\hyperref[corr:#1]{Corollary~\ref*{corr:#1}}}
\newcommand{\fig}[1]{\hyperref[fig:#1]{~\ref*{fig:#1}}}
\newcommand{\deff}[1]{\hyperref[deff:#1]{~\ref*{deff:#1}}}

\newcommand{\C}{\mathcal{C}}

\newcommand{\T}{\mathcal{T}}
\newcommand{\D}{\mathcal{D}}
\newcommand{\I}{\mathcal{I}}
\newcommand{\F}{\mathcal{F}}

\renewcommand{\H}{\mathcal{H}}

\newcommand{\B}{\mathcal{B}}

\renewcommand{\S}{\mathcal{S}}

\DeclareMathOperator{\conv}{conv}

\DeclareMathOperator{\cl}{cl}

\renewcommand{\succ}{\mathrm{succ}}

\newcommand{\RR}{\mathbb{R}}
\newcommand{\CC}{\mathbb{C}}

\DeclareMathOperator{\Tr}{Tr}

\newcommand{\id}{\mathbbm{1}}

\let\mathscr\relax
\DeclareFontFamily{U}{mathc}{}
\DeclareFontShape{U}{mathc}{m}{it}%
{<->s*[1.03] mathc10}{}
\DeclareMathAlphabet{\mathscr}{U}{mathc}{m}{it}

\newcommand{\ketbra}[2]{|{#1}\rangle\!\langle{#2}|}

\newcommand{\ba}{\begin{eqnarray}}
\newcommand{\ea}{\end{eqnarray}}
\newcommand{\bann}{\begin{eqnarray*}}
\newcommand{\eann}{\end{eqnarray*}}
\newcommand{\bal}{\begin{equation}\begin{aligned}}
\newcommand{\eal}{\end{aligned}\end{equation}}

\newcommand{\txb}[1]{{\color{blue!85!black} #1}}

\newcolumntype{L}[1]{>{\raggedright}p{#1}}
\newcolumntype{C}[1]{>{\centering}p{#1}}
\newcolumntype{R}[1]{>{\raggedleft}p{#1}}
\newcolumntype{D}{>{\centering\arraybackslash}X}

\newcommand{\norm}[1]{\left\|#1\right\|}

\let\C\C

\newcommand{\proj}[1]{\ket{#1}\!\bra{#1}}

\newcommand{\sbar}{\;\rule{0pt}{9.5pt}\right|\;}

\newcommand{\lset}{\left\{\left.}
\newcommand{\rset}{\right\}}

\let\txb\relax

\begin{document}

\title{Operational quantification of continuous-variable quantum resources}

\author{Bartosz Regula}
\email{bartosz.regula@gmail.com}
\affiliation{School of Physical and Mathematical Sciences, Nanyang Technological University, 637371, Singapore}
\author{Ludovico Lami}
\email{ludovico.lami@gmail.com}
\affiliation{Institut f\"ur Theoretische Physik und IQST, Universit\"at Ulm, Albert-Einstein-Allee 11, D-89069 Ulm, Germany}
\author{Giovanni Ferrari}
\affiliation{Dipartimento di Fisica e Astronomia Galileo Galilei,
Universit\`a degli studi di Padova, via Marzolo 8, 35131 Padova, Italy}
\affiliation{Institut f\"ur Theoretische Physik und IQST, Universit\"at Ulm, Albert-Einstein-Allee 11, D-89069 Ulm, Germany}
\author{Ryuji Takagi}
\affiliation{Center for Theoretical Physics and Department of Physics, Massachusetts Institute of Technology, Cambridge, Massachusetts 02139, USA}
\affiliation{School of Physical and Mathematical Sciences, Nanyang Technological University, 637371, Singapore}

\begin{abstract}%
The diverse range of resources which underlie the utility of quantum states in practical tasks motivates the development of universally applicable methods to measure and compare resources of different types. However, many of such approaches were hitherto limited to the finite-dimensional setting or were not connected with operational tasks.
We overcome this by introducing a general method of quantifying resources for continuous-variable quantum systems based on the robustness measure, applicable to a plethora of physically relevant resources such as optical nonclassicality, entanglement, genuine non-Gaussianity, and coherence. We demonstrate in particular that the measure has a direct operational interpretation as the advantage enabled by a given state in a class of channel discrimination tasks.  We show that the robustness constitutes a well-behaved, bona fide resource quantifier in any convex resource theory, contrary to a related negativity-based measure known as the standard robustness.
Furthermore, we show the robustness to be directly observable --- it can be computed as the expectation value of a single witness operator --- and establish general methods for evaluating the measure. Explicitly applying our results to the relevant resources, we demonstrate the exact computability of the robustness for several classes of states.
\end{abstract}

\maketitle


As quantum technologies begin to outperform classical ones in a number of practical applications~\cite{dowling_2003,preskill_2018}, it becomes crucial to precisely and efficiently characterize the advantages enabled by quantum mechanics. Depending on the particular task, different properties of quantum systems can be understood as the source of quantum advantages --- e.g., nonclassicality in quantum optics~\cite{braunstein_2005}, entanglement in communication scenarios~\cite{horodecki_2009}, non-Gaussianity in quantum computation~\cite{lloyd_1999,knill_2001,niset_2009}. This motivates a unified description of all such phenomena, allowing for the development of broadly applicable methods to characterize and quantify the various resources. 
The framework of \textit{quantum resource theories} was thus conceived to understand such physical properties within a common formalism~\cite{horodecki_2012,delrio_2015,coecke_2016,chitambar_2019}, which has led to many developments in the understanding of general classes of resources~\cite{brandao_2015,gour_2017,liu_2017,anshu_2018-1,regula_2018,lami_2018,takagi_2019-2,uola_2019-1,takagi_2019,sparaciari_2020,liu_2019,fang_2019-1,takagi_2020-1,regula_2020,kuroiwa_2020,seddon_2020,ducuara_2019,uola_2019-2}.

Although successful in describing finite-dimensional quantum theory, the commonly employed tools of quantum resource theories do not readily generalize to the infinite-dimensional setting. 
Exceptions to this rule typically only pertain to severely restricted frameworks such as the Gaussian one~\cite{lami_2018,lami_2020,jee_2020}. 
Such limitations make the methods inapplicable to general continuous-variable quantum systems, which are the cornerstone of many quantum technologies of fundamental importance~\cite{lloyd_1999,gisin_2007,braunstein_2005,giovannetti_2011}. 
This necessitated the development of resource-specific and mutually incompatible approaches to continuous-variable resources~\cite{hillery_1987,lee_1991,kenfack_2004,asboth_2005,vogel_2014,yadin_2018,kwon_2019,eisert_2002,adesso_2007-2,genoni_2010,idel_2016,zhang_2016,takagi_2018,albarelli_2018,narasimhachar_2019}, which obscures the connections and common features between resources of different types. In particular, one of the key applications of resource-theoretic concepts is to quantify the advantages that a resource can provide in practical tasks, but many resource measures defined in an abstract or ad-hoc manner lack such an operational meaning.

In this work, we address the need for a general approach to continuous-variable resource quantification by introducing the \textit{robustness} as a universal and operationally relevant measure. Inspired by a measure of entanglement~\cite{vidal_1999} which later found use in a range of discrete-variable settings~\cite{harrow_2003,napoli_2016,regula_2018,anshu_2018-1,takagi_2019-2,takagi_2019,liu_2019,regula_2020,seddon_2020}, the application of this quantifier to infinite-dimensional resources was hindered by the technical issues associated with infinite-dimensional spaces. We introduce an extension of the robustness to the continuous-variable setting, showing that it yields a valid and faithful resource monotone in any infinite-dimensional convex resource theory. Crucially, we show that the robustness exactly quantifies the advantage enabled by a given resource state in a class of channel discrimination tasks, thus endowing the quantifier with a direct operational interpretation. We establish accessible bounds and expressions for the robustness, in particular showing that it can always be evaluated by measuring a suitably chosen quantum observable. We compare the robustness to another measure commonly used in finite-dimensional theories, the so-called standard robustness~\cite{vidal_1999}, and show that the latter often fails to be a meaningful resource quantifier in infinite dimensions --- notably, for entanglement and nonclassicality theories --- highlighting our measure as a well-behaved continuous-variable monotone. To demonstrate the versatility of our framework and connect the results with physically relevant resources, we consider applications of our results to the resource theories of nonclassicality, entanglement, and coherence. We evaluate the robustness exactly for several classes of states in these theories, including Fock states and squeezed states as representative nonclassical states, and all pure entangled states.

We present a self-contained discussion of our methods and results below. The full technical details and additional developments are deferred to the companion paper~\cite{our_companion}, where we consider the problem of quantifying infinite-dimensional resources from a broader perspective of general probabilistic theories, extending the concepts discussed herein.


\textbf{\textit{General resource theories}}.--- %
The setting of our work will be an infinite-dimensional separable Hilbert space $\mathcal{H}$. We use $\B(\H)$ to denote the space of bounded linear operators on $\H$, and $\D(\H)$ for the set of density operators.
When discussing sequences of states or operators, we use the topology induced by the trace norm $\norm{\cdot}_{1}$.

A resource theory is a general framework \txb{describing} the manipulation of quantum states under some physically-motivated restrictions on the allowed operations~\cite{chitambar_2019}. In such a setting, only the states and channels from a certain set --- termed \emph{free} --- are freely available as they carry no resource, while states and operations outside of the designated set
are resourceful and thus costly to use. The paradigmatic example of a resource theory is entanglement~\cite{horodecki_2009}, where separable states together with local operations and classical communication (LOCC) are free.

We will now consider a general resource-theoretic setting. So as to ensure the broadest applicability of our results, we will only make two intuitive assumptions about the set of free states $\F \subseteq \D(\H)$. First, we take $\F$ to be closed, \txb{that is,}
a sequence of free states cannot converge to a resourceful state, since the outcomes of experiments should be consistent under limits. Secondly, we assume $\F$ to be convex, \txb{i.e., no resource can be generated by probabilistically mixing free states.} From an operational point of view as a resource, even if a set of interest is not convex (e.g., the set of all Gaussian states), the convex combinations of such states can often also be taken to be free~\cite{takagi_2018,albarelli_2018}, making convexity a natural property of operational quantum resources. Thus, all of the fundamental continuous-variable resource theories such as entanglement, nonclassicality, genuine (convex) non-Gaussianity, and coherence can be described in our framework.

Similarly, we only make the weakest possible assumption on the allowed set of free operations; namely, that a free state remains free under the action of a free transformation. All choices of free operations --- be it classical processes~\cite{gehrke_2012,rahimi-keshari_2013} or linear optical transformations~\cite{yadin_2018,kwon_2019} in the theory of nonclassicality, LOCC or non-entangling operations~\cite{brandao_2011} in entanglement theory~\cite{horodecki_2009}, or any other physical class of resource transformations --- are thus encompassed in this framework.


\textbf{\textit{Operational resource quantifier}}.--- %
A resource theory, as defined in the previous section, is a purely abstract concept. A natural question then arises of whether any theory defined in this way truly represents a ``resource'' --- that is, does any state $\rho \notin \F$ provide a practical advantage over the resourceless states in $\F$? It was shown in~\cite{takagi_2019-2} that it is indeed the case, and all resources can be useful in channel discrimination tasks. Such tasks are fundamental to the operational description of quantum states~\cite{kitaev_1997,wilde_2017} and underlie practical applications such as quantum illumination~\cite{lloyd_2008,tan_2008} and sensing~\cite{pirandola_2018}. However, the quantitative methods used to study discrimination problems~\cite{piani_2015,piani_2016,takagi_2019-2} are limited to finite dimensions, and it is not \textit{a priori} clear how to measure the advantages provided by continuous-variable resources. Here, we will introduce an operational resource quantifier which precisely benchmarks the maximal advantage facilitated by a given resource in discrimination tasks.

In order to measure and compare the resource content of states, we employ the concept of robustness measures~\cite{vidal_1999}, which quantify how much noise is required to destroy the resources contained in a given state. Specifically, in finite-dimensional theories, the \textit{(generalized) robustness} is defined as
\begin{equation}\begin{aligned}\label{eq:rob_nonlsc_definition}
	\R_\F(\rho) \coloneqq \inf_{\substack{\lambda \in \RR_+
	}} \lset 1+\lambda \sbar \frac{\rho + \lambda \tau}{1 + \lambda} \in \F, \; \tau \in \D(\H) \rset
\end{aligned}\end{equation}
which corresponds to the least coefficient such that the mixture of $\rho$ with a noise state $\tau$ becomes a free state. We note that this definition differs by a term $+1$ from the original notation of~\cite{vidal_1999}, a choice we made for mathematical convenience.
In order to ensure full generality in infinite dimensions, we will additionally allow optimization over all sequences of operators $\{\xi_k\}_k \in \B(\H)$ which converge to the given state $\rho$. We thus take our definition of the robustness to be
\begin{equation}\begin{aligned}\label{eq:rob_definition}
	\Rl_\F(\rho) \coloneqq \inf_{\substack{\lambda \in \RR_+
	}} \bigg\{ 1+\lambda \;\bigg|\;& \frac{\xi_k + \lambda \tau_k}{1 + \lambda} \in \F, \; \tau_k \in \D(\H),\\
	& \{\xi_k\}_k \to \rho \,\bigg\}.
\end{aligned}\end{equation}
Despite the seemingly more complicated form, we will shortly show that there are many efficient ways to bound and compute this measure.
In many of the practically relevant resource theories such as nonclassicality and entanglement, we show that $\Rl_\F(\rho)=\R_\F(\rho)$, meaning that the optimization over sequences is not necessary.
We invite the interested reader to~\cite{our_companion} for a discussion of the technical issues concerning the definition of robustness in infinite dimensions.

In the task of channel discrimination, a channel is randomly selected from an ensemble $\{p_i, \Lambda_i\}$ of quantum channels $\Lambda_i$ with corresponding probabilities $p_i$. After sending a chosen state $\rho$ through the channel, the player is then tasked with determining which of the channels $\{\Lambda_i\}$ was applied by performing a measurement of the output state. By the Born rule, the average probability of successful discrimination with a positive operator-valued measure (POVM) $\{M_i\}$ is then given by $p_{\succ}(\rho, \{p_i, \Lambda_i\}, \{M_i\}) = \sum_i p_i \Tr[ M_i \Lambda_i(\rho) ]$. For simplicity, we will use $\T = \{ \{p_i, \Lambda_i\}, \{M_i\} \}$ to denote a given discrimination task.

In order to directly quantify the advantage provided by a given state $\rho$, we then ask: all else being equal, how much better can the player perform in the given task by using the state $\rho$ instead of a free state $\sigma \in \F$? 
We show that the maximal such advantage is given precisely by the robustness.

\begin{thm}\label{thm:robustness_operational}
For any state $\rho \in \D(\H)$, it holds that
\begin{equation}\begin{aligned}
	\sup_{\T} \frac{p_{\succ}(\rho, \T)}{\sup_{\sigma \in \F} p_{\succ}(\sigma, \T)} = \Rl_\F(\rho),
\end{aligned}\end{equation}
where the maximization is over all discrimination tasks $\T = \{ \{p_i, \Lambda_i\}, \{M_i\} \}$, i.e., all channel ensembles $\{p_i, \Lambda_i\}_{i=1}^n$ and POVMs $\{M_i\}_{i=1}^n$ with $n$ arbitrary.

As long as $\Rl_\F(\rho) < \infty$, there exists a discrimination task with $n=2$ which achieves this supremum.
\end{thm}
The proof employs the theory of optimization in Banach spaces~\cite{kretschmer_1961,ponstein_2004} to relate the robustness with an optimization of quantum observables, establishing a convex duality relation which provides a new extension of approaches used in finite-dimensional theories.

We have thus shown a \txb{direct operational meaning} for the robustness $\Rl_\F$ in any convex quantum resource theory. \txb{The implications of Theorem~\ref{thm:robustness_operational} for particular tasks, such as phase discrimination~\cite{napoli_2016}, were already known. The merit of our result is to demonstrate the generality of this approach, which carries over to continuous-variable resource theories.} Further extensions and applications of Theorem~\ref{thm:robustness_operational} are described in~\cite{our_companion}.

\textbf{\textit{Robustness as a resource monotone}}.--- %
For a function $M : \D(\H) \to \RR_+ \cup \{\infty\}$ to be considered a meaningful resource quantifier, it is generally required to satisfy several properties~\cite{chitambar_2019}. The most important is \emph{monotonicity}, that is, $M(\Phi(\rho)) \leq M(\rho)$ under the action of a free operation $\Phi$.
A stronger type of monotonicity is often imposed, ensuring that the measure cannot increase on average in probabilistic transformations~\cite{vidal_2000}. Another feature is the \emph{faithfulness} of the measure, that is, the property that $M(\rho)$ achieves its minimal value if and only if $\rho \in \F$, which is necessary to precisely delineate the resource character of quantum states in consideration. 

As our next result, we then show that the robustness is a valid resource monotone in any convex resource theory. To demonstrate the strongest type of monotonicity, we model probabilistic transformations in the general formalism of quantum instruments~\cite{davies_1970}, i.e., as a collection of completely positive maps (subchannels) $\{\Phi_i\}_{i}$ which map $\rho$ to $\Phi_i(\rho)$ with corresponding probability $p_i = \Tr \Phi_i(\rho)$ and the overall transformation $\sum_i \Phi_i$ is trace preserving. A free instrument then satisfies $\Phi_i(\sigma) \propto \sigma' \in \F$ for all $\sigma \in \F$.
\begin{thm}\label{thm:rob_measure_properties}
The robustness $\Rl_\F$ is: (i) convex; (ii) faithful, i.e., $\Rl_\F(\rho) = 1$ if and only if $\rho \in \F$; (iii) monotonic on average under probabilistic free operations --- that is, for any free instrument $\{\Phi_i\}_{i}$, it holds that $\displaystyle  \Rl_\F(\rho) \geq \sum_i p_i \Rl_\F\left(\frac{\Phi_i(\rho)}{p_i}\right)$.
\end{thm}

Together with Thm.~\ref{thm:robustness_operational}, the faithfulness of the robustness implies that any state $\rho \notin \F$ provides an advantage over all $\sigma \in \F$ in a practical task of channel discrimination, recovering a result of \cite{takagi_2019-2}. Such advantages are often non-trivial to show, owing to the existence of so-called bound resources~\cite{horodecki_2001-1,horodecki_2000,campbell_2010,mari_2012,veitch_2013,takagi_2018} which provide no advantage in certain tasks. Although measures such as the entanglement negativity~\cite{vidal_2002} or Wigner negativity~\cite{kenfack_2004} can be easier to compute than $\Rl_\F$, they fail to detect the resources of such bound states.

If the noise state $\tau$ in the definition of $\R_\F$ (Eq.~\eqref{eq:rob_nonlsc_definition}) is constrained to be a free state, it gives a related resource measure known as the \textit{standard robustness}~\cite{vidal_1999,Note0}:
\begin{equation}\begin{aligned}
	\Rs_\F(\rho) \coloneqq& \inf \lset 1+\lambda \sbar \frac{\rho + \lambda \sigma}{1 + \lambda} \in \F, \; \sigma \in \F \rset,
\end{aligned}\end{equation}
which corresponds to the negative part of a linear decomposition of $\rho$ into free states, generalizing the notion of negativity~\cite{vidal_2002}. A problem concerning the quantification of continuous-variable resources is that the existence of infinitely resourceful states with respect to certain tasks is a physical possibility~\cite{eisert_2002,keyl_2003}. 
However, a meaningful quantifier should not yield an infinite value for states which are not infinitely resourceful. We will shortly see that the standard robustness often suffers from this problem --- that is,  in several resource theories there exist classes of states for which $\Rl_\F(\rho) < \infty = \Rs_\F(\rho)$, making $\Rs_\F$ useless for benchmarking the resources of such states. The generalized robustness $\Rl_\F$ is therefore more suited to be a universal resource quantifier, which is why we do not discuss the quantity $\Rs_\F$ in detail.

\textbf{\textit{Evaluating the robustness}}.--- %
It will be useful to express the robustness in its dual form. Although frequently employed in finite-dimensional resource theories~\cite{brandao_2005,regula_2018,takagi_2019-2,uola_2019-1,takagi_2019}, convex duality has seen few applications in infinite-dimensional quantum information. 
We establish a general expression for the robustness, providing a non-trivial extension of methods and results that have only been obtained in finite-dimensional cases~\cite{brandao_2005,regula_2018}.
\begin{thm}\label{thm:rob_dual_form}
For any state, the robustness can be computed as
\begin{equation}\begin{aligned}\label{eq:rob_dual_form}
	\Rl_\F(\rho) \!=\! \sup \lset \Tr W \rho \sbar\! W\! \in\! \B(\H),\, W \!\geq\! 0,\, \Tr W \sigma \!\leq\! 1 \ \forall \sigma\! \in\! \F \rset ,
\end{aligned}\end{equation}
and the supremum is achieved as long as it is finite.
\end{thm}
The result of Thm.~\ref{thm:rob_dual_form} shows in particular that the robustness is directly observable in {experiments} without the need for state tomography, as it can be evaluated by computing the expectation value of a suitably chosen quantum observable. Indeed, we can understand the feasible solutions $W$ to Eq.~\eqref{eq:rob_dual_form} as resource witnesses~\cite{terhal_2002} which can detect and certify the resourcefulness of states --- for any $\rho \notin \F$, there exists a witness such that $\Tr(W\sigma) \leq 1 \; \forall \sigma \in \F$ but $\Tr(W\rho) >1$. 
Any resource witness provides an accessible lower bound for the robustness, and crucially, we can always find a witness which achieves the optimum.

We obtain several general bounds to the robustness~\cite{our_companion}, a particularly useful one being as follows.
\begin{lem}\label{lem:rob_pure_bounds}
	For any pure state, we have
\begin{equation}
\inf_{\sigma \in \F} \,\braket{\psi|\sigma|\psi}^{\!-1} \leq \displaystyle \Rl_\F(\psi) \leq \inf_{\sigma\in \F} \braket{\psi|\sigma^{-1}|\psi}.
\label{eq:rob_pure}
\end{equation}
\end{lem}

\noindent Here, $\braket{\psi|\sigma^{-1}|\psi} = \left\|\sigma^{-1/2} \ket{\psi}\right\|^2$, and the infimum is restricted to states such that $\ket{\psi}$ lies in the domain of $\sigma^{-1/2}$~\cite{kraus_1973}. 
Lemma~\ref{lem:rob_pure_bounds} gives 
\txb{computationally useful methods to bound the robustness in both directions.}
We proceed by explicitly applying our results in important resource theories. In all of the examples considered below, we can show that $\Rl_\F(\rho) = \R_\F(\rho)$~\cite{our_companion}.


\textbf{\textit{Nonclassicality}}.--- %
Nonclassicality is a fundamental continuous-variable resource concerned with exploiting the truly quantum properties of light~\cite{glauber_1963,sudarshan_1963,braunstein_2005}. 
Its quantification attracted significant attention~\cite{hillery_1987,lee_1991,kenfack_2004,asboth_2005,gehrke_2012,vogel_2014,nair_2017,tan_2017,yadin_2018,kwon_2019,ge_2020,tan_2020,relentNC} and recently it was formalized as a resource theory~\cite{yadin_2018,kwon_2019}. Let us then consider the quantum theory of a single harmonic oscillator 
(an extension to multiple modes being straightforward).
The free states here are the so-called classical states~\cite{glauber_1963,sudarshan_1963}: defining the coherent states $\ket{\alpha}\coloneqq e^{-|\alpha|^2/2}\sum_{n=0}^\infty \frac{\alpha^n}{\sqrt{n!}} \ket{n}$ where $\ket{n}$ denotes the $n$th Fock state, we have that $\F = \C \coloneqq \cl \conv \lset \proj{\alpha} \sbar \alpha\in \CC \rset$ where $\cl \conv$ denotes closed convex hull, and any state outside of this set is nonclassical~\cite{Bach1986}. 
Notable examples of nonclassical states include the Fock states themselves, the squeezed states $\ket{\zeta_r} = e^{r (a^2 - a^{\dagger 2}) / 2} \ket{0}$, and the Schr\"odinger cat states $\ket{\alpha_\pm} \propto \ket{\alpha} \pm \ket{-\alpha}$.

We will first show that the standard robustness of nonclassicality, $\Rs_\C$, is infinite for a large class of pure states in this resource theory, effectively rendering the measure useless in discerning the resourcefulness of different states~\footnote{We remark that the unboundedness of $\Rs_\C$ for specific states was previously mentioned in Ref.~\cite{tan_2020} without proof.}. To this end, we establish a lower bound as $\Rs_\C(\rho) \geq \frac{1}{2}\left(\sup_{\alpha\in \CC} \left|\chi_1^\rho(\alpha)\right|+1\right)$, where $\chi_1^\rho(\alpha)\coloneqq e^{|\alpha|^2/2}\Tr\left[\rho e^{\alpha a^\dagger - \alpha^* a}\right]$ denotes the normally-ordered characteristic function. Hence, any state with unbounded $\chi_1^\rho$ necessarily has infinite standard robustness --- we show that this comprises most physically accessible classes of nonclassical states including finite superpositions of Fock states, squeezed states, cat states, and nonclassical Gaussian states. 

Conversely, we will show that the generalized robustness $\Rl_\C$ is a well-behaved quantifier. For Fock states $\ket{n}$, we find that the lower and upper bounds from Lemma~\ref{lem:rob_pure_bounds} coincide by employing a phase-randomized coherent state $\sigma_n = \frac{1}{2\pi} \int_0^{2\pi} \proj{\sqrt{n} e^{i\theta}} \mathrm{d}\theta$, giving
\begin{equation}\begin{aligned}
	\Rl_\C(\proj{n}) = e^{n} \frac{n!}{n^n}.
\end{aligned}\end{equation}
For squeezed states $\ket{\zeta_r}$, an application of Lemma~\ref{lem:rob_pure_bounds} with a construction of an ansatz based on the thermal state $\tau_N\coloneqq \frac{1}{N+1} \sum_{n=0}^\infty \left( \frac{N}{N+1} \right)^n \ketbra{n}{n}$ gives
\begin{equation}\begin{aligned}
	\Rl_\C(\zeta_r) = e^r
\end{aligned}\end{equation}
for any $r \geq 0$. For cat states $\ket{\alpha_\pm}$ with $\alpha \geq 0$, Lemma~\ref{lem:rob_pure_bounds} gives an upper bound of $ \Rl_\C(\alpha_\pm) \leq 2 \left(1\pm e^{-2\alpha^2}\right)^{-1}$, which can be numerically verified to be tight for $\ket{\alpha_+}$~\cite{our_companion}. The corresponding lower bound gives $\Rl_\C(\alpha_\pm) \to 2$ as $\alpha$ increases. %
A similar approach can be applied to give bounds for single-photon-added or single-photon-subtracted squeezed states. 
We thus see that the robustness $\Rl_\C$ is not only finite, but can in fact be efficiently computed in many relevant instances. 


\textbf{\textit{Entanglement}}.--- %
Entanglement underlies many of the nonlocal features of quantum mechanics and has found use in a variety of continuous-variable settings~\cite{braunstein_2005,horodecki_2009,adesso_2007-2,horodecki_2009}. The resource theory of entanglement is defined in a bipartite Hilbert space $\H = \H_A \otimes \H_B$, 
where the free states are the separable states $\F = \S \coloneqq \cl \conv \lset \proj\psi \sbar \ket\psi = \ket{\psi_A} \otimes \ket{\psi_B} \rset$~\cite{werner_1989}.

We will again show that there exists an example of a state such that the standard robustness $\Rs_\S$ is infinite, while the generalized robustness $\Rl_\S$ remains a well-behaved quantifier. To our knowledge, no explicit state having different values of the two measures has been presented before, even in finite dimensions. 
We will establish an even stronger result by showing that the entanglement negativity --- a quantifier commonly employed in practical settings --- is also infinite for this state. Recall that the negativity is defined as $\frac{1}{2}\left(\norm{\rho^\Gamma}_{1} - 1\right)$~\cite{vidal_2002}, where $\rho^\Gamma$ is a partial transpose of $\rho$. To construct our ansatz, we employ the Hilbert operator $H_{-1} \in \B(\H_A)$~\cite{WeylPhD,Schur1911,INEQUALITIES} whose matrix elements $(H_{-1})_{n,m}$ ($n,m\geq 1$) are given by $0$ if $n=m$ or $\frac{1}{n-m}$ otherwise, and define the single-party states $\omega_\pm \coloneqq \frac{1}{c} D\left(\id \pm \frac{i}{\pi} H_{-1}\right) D$, where $D \coloneqq \sum_{n=1}^\infty \frac{1}{\sqrt{n}\ln(n+1)} \proj{n}$ and $c$ is a normalization constant. 
Constructing now the bipartite states $\rho_\pm \coloneqq \sum_{n,m=1}^\infty (\omega_\pm)_{n,m} \ketbra{nn}{mm}$, we compute $\norm{\rho_\pm^\Gamma}_{1} = \infty$~\cite{our_companion,INEQUALITIES}. Since the negativity lower bounds the standard robustness $\Rs_\S$~\cite{vidal_2002}, we get $\Rs_\S (\rho_\pm) = \infty$. However, $\frac{1}{2}(\rho_+ + \rho_-)$ is a separable state, hence $\Rl_\S(\rho_\pm) \leq 2$.

We now consider the case of general pure states. Applying the dual characterization in Thm.~\ref{thm:rob_dual_form} and adapting an argument of Ref.~\cite{vidal_1999}, we show that the finite-dimensional formula for the robustness~\cite{vidal_1999,harrow_2003,steiner_2003} generalizes to infinite dimensions. Specifically, it holds that $\Rl_\S(\psi) = \left(\sum_{n=1}^\infty \mu_n\right)^2$, 
where $ \ket\psi = \sum_{n=1}^{\infty} \mu_n \ket{nn}$ (with $\mu_n\geq 0$) 
is the Schmidt decomposition of the given state. This shows that the robustness of a pure state is finite iff the sum of the Schmidt coefficients converges. 
Interestingly, for a two-mode squeezed vacuum state $\ket{\nu_r}$ we get $\Rl_\S(\nu_r) = e^r$, which equals the nonclassicality of the single-mode squeezed state $\ket{\zeta_r}$.

\textbf{\textit{Coherence}}.--- %
The resource theory of quantum coherence (not to be confused with coherent states in the theory of nonclassicality) is concerned with quantifying superposition in an orthonormal basis of the Hilbert space~\cite{baumgratz_2014,streltsov_2017,zhang_2016}, e.g., the Fock basis $\{\ket{n}\}_{n=1}^\infty$. All diagonal states $\F = \I \coloneqq \cl \conv \{ \proj{n} \}_{n=1}^\infty$ are considered free. 

An extension of our argument for entanglement theory can be used to show that, for any pure state $\ket\psi = \sum_{n=1}^\infty \psi_n \ket{n}$, the robustness $\Rl_\I(\psi)$ equals the $\ell^1$ norm $\norm{\psi}_{\ell^1} = (\sum_{n=1}^\infty |\psi_n|)^2$. However, $\Rl_\I$ is in general smaller than the $\ell^1$ norm, and this difference can be arbitrarily large --- for the states $\omega_\pm$ that we considered in our discussion of entanglement, it holds that $\Rl_\I(\omega_\pm) \leq 2$ but $\norm{\omega_\pm}_{\ell^1} = \infty$. This once again shows that the robustness can be a more well-behaved quantifier than other common measures.

\textbf{\textit{Other resources}}.--- %
We stress that our results are readily applicable to any other convex resource theory. In the companion paper~\cite{our_companion}, we present an explicit application of our methods to genuine non-Gaussianity~\cite{takagi_2018,albarelli_2018}. We further provide additional results and extensions, including general criteria for the strong duality $\R_\F(\rho) = \Rl_\F(\rho)$ to hold, as well as a connection between the robustness and a class of norm-based measures~\cite{arveson_2009} which generalizes the relation with the $\ell^1$ norm~\cite{rudolph_2001,harrow_2003,steiner_2003,piani_2016,regula_2018,seddon_2020}.


\textbf{\textit{Discussion}}.--- %
We introduced the generalized robustness as an operational measure of general convex quantum resources in infinite dimensions. We showed in particular that it exactly quantifies the advantage provided by a given resource state in a class of operational discrimination tasks, directly relating the measure with the practical exploitation of quantum resources. We established methods for lower and upper bounding the robustness, showing them to be tight in many cases and in general providing theoretically and experimentally accessible ways of evaluating the measure. Finally, we showcased the broad applicability of the quantifier by explicitly applying it to characterize the theories of nonclassicality, entanglement, and coherence. The results provide accessible methods for benchmarking the diverse resources which underlie technological applications, facilitating their operational description.

Our work opens an avenue for the study of operational aspects of general continuous-variable quantum resources. Furthermore, the applications of a similar formalism to finite-dimensional quantum channels~\cite{diaz_2018-2,takagi_2019,uola_2019,yuan_2020,takagi_2020,takagi_2020_optimal, liu_2019_channel} and measurements~\cite{skrzypczyk_2019-1,uola_2019-1,takagi_2019,oszmaniec_2019-1} suggest that analogous extensions could be possible in infinite dimensions, with the first steps already taken in~\cite{kuramochi_2020}.


\begin{acknowledgments}
\textit{Note added}.--- After the completion of this work, we became aware of Ref.~\cite{haapasalo_2020}, where a related robustness-based approach to continuous-variable resource quantification is considered using a complementary set of methods which rely on the construction of finite-dimensional approximations to infinite-dimensional measures.

B.R. is supported by the Presidential Postdoctoral Fellowship from Nanyang Technological University, Singapore. L.L. is supported by the ERC Synergy Grant BIOQ (grant no. 319130) and by the Alexander von Humboldt Foundation. G.F. acknowledges the support received from the EU through the ERASMUS+ Traineeship program and from the Scuola Galileiana di Studi Superiori. R.T. acknowledges the support of NSF, ARO, IARPA, AFOSR, the Takenaka Scholarship Foundation, and the National Research Foundation (NRF) Singapore, under its NRFF Fellow programme (Award No. NRF-NRFF2016-02) and the Singapore Ministry of Education Tier 1 Grant 2019-T1-002-015. Any opinions, findings and conclusions or recommendations expressed in this material are those of the author(s) and do not reflect the views of National Research Foundation, Singapore.
\end{acknowledgments}

\bibliographystyle{apsrev4-1a}
\bibliography{main}

\end{document}